\documentclass{article}
\usepackage[english]{babel}
\usepackage{amsmath,amssymb,tabularx,bm,array,booktabs}
\usepackage{xcolor,slashed,epsfig,cite}
\usepackage{verbatim,slashed,ulem,geometry,setspace,algorithm}
\usepackage{algorithmic,mathrsfs,braket,multirow,dcolumn,subfigure}
\usepackage{graphicx}
\usepackage[colorlinks=true,citecolor=blue,urlcolor=cyan,filecolor=magenta,linkcolor=red,frenchlinks=true,bookmarks]{hyperref}
\allowdisplaybreaks[4]
\renewcommand\sout{\bgroup \color{red}  \ULdepth=-.5ex \ULset}
\newenvironment{mysubeq}
{\begin{subequations}
}
{\end{subequations}}

\begin{document}

\title{Form factors of $\Omega^-$ in a covariant quark-diquark approach}

\author{Dongyan Fu$^{a,b,\thanks{fudongyan@ihep.ac.cn}}$,
JiaQi Wang$^{a,b,\thanks{jqwang@ihep.ac.cn}}$,
and Yubing Dong$^{a,b,\thanks{dongyb@ihep.ac.cn}}$\\
Institute of High Energy Physics, Chinese Academy
of Sciences, Beijing 100049, China$^{a}$\\
School of Physical Sciences, University of Chinese
Academy of Sciences, Beijing 101408, China$^{b}$\\
}
\maketitle
\begin{abstract}
The electromagnetic and gravitational form factors of $\Omega^-$, a spin-3/2 hyperon 
composed of three $s$ quarks, are calculated by using a covariant quark-diquark approach. 
The model parameters are determined by fitting to the form factors of the 
lattice QCD calculations. Our obtained electromagnetic radii, magnetic moment, and 
electric-quadrupole moment are in agreement with  the experimental measurements and some 
other model calculations. Furthermore, the mass and spin distributions of $\Omega^-$ 
from the gravitational form factors are also displayed. It is found that the mass 
radius is smaller than its electromagnetic ones. Finally, the interpretations 
of the energy density and momentum current distribution are also discussed.
\end{abstract}

\section{Introduction}

\quad\quad Form factors (FFs), such as electromagnetic form factors (EMFFs) and gravitational form factors (GFFs), are the very important physical quantities that describe the internal structure of a hadron. They carry the fundamental and essential information, such as the distributions of the electric charge, magnetic moment, mass and spin. There have been many studies of EMFFs \cite{Gilman:2001yh,Gross:2002ge,Dong:2008mt,deMelo:2008rj,Cloet:2014rja,Sun:2016ncc} and  GFFs \cite{Polyakov:2002yz,Kumano:2017lhr,Polyakov:2018exb,Lorce:2018egm,
Polyakov:2018zvc}. In particular much work has been devoted to study the properties 
of the various low-spin hadrons, such as spin-0 ($\pi$  \cite{Hohler:1974eq,Maris:2000sk,Kumano:2017lhr}), spin-1/2 (nucleon \cite{Bincer:1959tz,Sugawara:1968ty,Burkert:2018bqq,Burkert:2023wzr}), and spin-1 ($\rho$ \cite{Sun:2017gtz} and 
deuteron  \cite{Berger:2001zb,Cano:2003ju,Dong:2013rk,Kim:2022wkc}). 
Experimentally, EMFFs can be detected from the processes driven by the electromagnetic interactions. The 
corresponding processes, such as the hadron production processes from $e^+ e^-$ annihilation  \cite{CLEO:2005tiu,Seth:2012nn,Dobbs:2014ifa} and the inverse processes~\cite{Andreotti:2003bt}, are accessible. However, the direct detection of 
GFFs is not realistic due to the weak gravitational interaction. Fortunately, they 
can be obtained via the generalized parton distributions (GPDs) \cite{Ji:1998pc,Goeke:2001tz,Burkardt:2002hr,Diehl:2003ny,Belitsky:2005qn}, and GPDs 
can be extracted from deeply virtual Compton scattering (DVCS) by using sum rules, from vector-meson electro-production processes, and from generalized distribution amplitudes (GDAs) \cite{Kumano:2017lhr}.

As the total spin of the system increases, there are more FFs, such as electric-quadrupole, magnetic-octupole, energy-quadrupole, and angular 
momentum-octupole form factors for a spin-3/2 particle. Although some of the spin-3/2 
particles have been discussed and studied \cite{Schlumpf:1993rm,Lee:1997jk,Dey:1999fi,
Buchmann:2002et,Hong:2007pr,Kim:2019gka,Kim:2020lrs}, their detailed information 
is still lacking compared to those of the low-spin particles. $\Delta$ resonance 
is the typical spin-3/2 particle which has been usually 
considered \cite{Aubin:2008qp,Alexandrou:2008bn,Kim:2020lrs,Kim:2023yhp}. Another 
typical spin-3/2 particle is $\Omega^-$ \cite{Aubin:2008qp,Dobbs:2014ifa,
Ramalho:2020laj}. However, most of the work, in the literature, only 
focus on its EMFFs.

Comparing the $\Omega^-$ hyperon with the $\Delta$ resonance, we see that the 
former has a 
longer lifetime $(c\tau=2.461~\text{cm})$ and it undergoes weak decays. Therefore, we 
expect that $\Omega^-$ is more realistic to be measured. On the one hand, the $\Omega^-$ 
form factors in the time-like region have been measured by the $e^+ e^- \rightarrow \Omega^- \bar{\Omega}^+$ at CLEO \cite{Dobbs:2014ifa}. Based on the $e^+ e^- \rightarrow B \bar{B}$ process, the facilities, such as BABAR \cite{BaBar:2005pon,BaBar:2007fsu}, BES III \cite{BESIII:2017hyw,BESIII:2020uqk,Yuan:2021yks}, CLEO \cite{Dobbs:2017hyd}, and PANDA \cite{PANDA:2016scz}, all have the chance to measure its structures by producing the secondary $\Omega^-$ beam. In addition, $\Omega^-$ event can also be produced in the inclusive reaction $p + Be \rightarrow \Omega + X$ \cite{E756:1996fgc}.
On the other hand, the more promising and reliable method to describe the FFs of 
$\Omega^-$ is the Lattice QCD (LQCD). Except for some LQCD calculations \cite{Leinweber:1992hy,Alexandrou:2010jv,Boinepalli:2009sq}, there are also some model calculations of FFs, such as the chiral constituent quark model ($\chi$CQM)  \cite{Wagner:2000ii,Broniowski:2008hx,Kim:2020nug}, the chiral perturbation theory ($\chi$PT)~\cite{Butler:1993ej,Li:2016ezv}, the $1/N_c$ expansion~\cite{Luty:1994ub,Buchmann:2018nmu}, the SU(2) Skyme model \cite{Kim:2012ts}, the bag model \cite{Neubelt:2019sou}, the QCD sum rule (QCDSR) \cite{Lee:1997jk,Azizi:2019ytx},  the relativistic quark model (RQM) \cite{Schlumpf:1993rm,Ramalho:2009gk}, the non-relativistic quark model (NRQM) \cite{Isgur:1981yz,Krivoruchenko:1991pm}, and so on. 

In this work, we give a study of the electromagnetic properties of $\Omega^-$ and its mechanical properties. It should be 
remained that prior to this work, we have carried out the calculations and analyses 
for the FFs of the $\Delta$ resonance and for its  generalized parton distributions 
in a covariant quark-diquark approach  \cite{Fu:2022rkn,Fu:2022bpf,Fu:2023dea}. 
Here the same approach is employed to study the FFs of the $\Omega^-$ hyperon~\cite{Cloet:2014rja,Fu:2022rkn}.  We know that $\Omega^-$ is composed of three $s$ quarks and it is convenient to consider the two $s$ quarks as a whole, i.e. as an axial-vector diquark. Thus, we may deal with an effective two-body system without 
losing the main internal structure information, and consequently, the final FFs can be obtained by summing the contributions of the quark and diquark. It should be addressed that the diquark structure can be explicitly considered by replacing the quark electromagnetic and energy momentum tensor (EMT) currents by the corresponding ones of the diquark.

This paper is organized as follows. In Sec.~\ref{sectionapproach}, FFs and the quark-diquark approach are briefly discussed. Our numerical results of EMFFs in comparison with the results of LQCD and our GFFs are given in Sec.~\ref{sectionresults}, where our obtained energy and angular momentum distributions and their representations in the coordinate space are also displayed. In addition, the quantities related to the "$D$-term", pressures, and shear forces are  discussed as well. Finally, section \ref{sectionsummary} is devoted to a summary.
    
\section{Form factors and quark-diquark approach}\label{sectionapproach}
\subsection{Form factors of the spin-3/2 system}

\quad\quad  In this work, the same approach as Ref.~\cite{Fu:2022rkn} is employed to study 
the FFs of the $\Omega^-$ hyperon. For a spin-3/2 particle, the matrix element of the 
electromagnetic current can be written, in terms of the form factors $F_{i,j}^{V,a}$ 
as \cite{Cotogno:2019vjb}
\begin{equation}\label{vectorcurrent}
    \begin{split}
    \left\langle p^\prime,\lambda^\prime \left| \hat{J}_a^{\mu}\left( 0 \right)
    \right| p,\lambda\right\rangle=&-\bar{u}_{\alpha'} \left( p',\lambda' \right)
    \biggl[ \frac{P^\mu}{M} \left( g^{\alpha' \alpha } F_{1,0}^{V,a} \left( t \right)
    -\frac{q ^{\alpha'} q ^\alpha}{2M^2} F_{1,1}^{V,a} \left( t \right)\right)\\
    &+\frac{i \sigma^{\mu \nu} q_\nu}{2M} \left( g^{\alpha' \alpha} F_{2,0}^{V,a} \left( t \right)
    -\frac{q ^{\alpha'} q ^\alpha}{2M^2}F_{2,1}^{V,a} \left( t \right)\right) \biggr]
    u_\alpha \left( p,\lambda \right),
    \end{split}
\end{equation}
where $u_\alpha \left( p,\lambda\right)$ is the Rarita-Schwinger spinor and the normalization is taken to be $\bar{u}_{\sigma'}(p) u_\sigma(p)=-2 M \delta_{\sigma' \sigma}$ with $M$ being the $\Omega^-$ mass. In Eq.~\eqref{vectorcurrent} the kinematical variables $P^\mu=(p^\mu + p'^\mu)/2$, $q^\mu=p'^\mu-p^\mu$, and $t = q^2$ are employed 
and $p \, (p')$ is the momentum of the initial (final) state. Moreover, the form factors $F_{i,j}^{V,a}$ are defined flavor by flavor and include the contribution of gluon in general. The total form factors $F_{i,j}^V=\underset{a}{\sum} F_{i,j}^{V,a}$ are obtained as the index $a$ runs from the quark to gluon. Here, since we only consider the 
constituent quark, the gluon contribution is simply and effectively included.

In our numerical calculation, the average of the initial and final momenta 
is defined as $P^\mu = (E, \bm{0})$ and the momentum transfer is 
$q^\mu = (0,\bm{q})$ by using the Breit frame. Thus, $t=q^2=-\bm{q}^2=4(M^2 - E^2)$. 
The EMFFs of the spin-3/2 particle can be further expressed in terms of the electromagnetic covariant vertex function coefficients $F_{i,j}^V$, where $i=1,2$ and $j=0,1$, as \cite{Nozawa:1990gt}
\begin{mysubeq}\label{EMFFsequation}
    \begin{align}
        G_{E0}\left( t \right)=&\left( 1+\frac{2}{3}\tau \right) [F_{2,0}^V(t) + (1+\tau)(F_{1,0}^V(t)-F_{2,0}^V(t))] +\frac{2}{3} \tau (1+\tau) [F_{2,1}^V(t) + (1+\tau)(F_{1,1}^V(t)-F_{2,1}^V(t))],
    \\
        G_{E2}\left( t \right)=& [F_{2,0}^V(t) + (1+\tau)(F_{1,0}^V(t)-F_{2,0}^V(t))] + (1+\tau) [F_{2,1}^V(t) + (1+\tau)(F_{1,1}^V(t)-F_{2,1}^V(t))],
    \\
        G_{M1}\left( t \right)=&\left(1+\frac{4}{5}\tau\right) F_{2,0}^V \left( t \right)
        +\frac{4}{5} \tau  (\tau +1) F_{2,1}^V \left( t \right),
    \\
        G_{M3}\left( t \right)=& F_{2,0}^V \left( t \right)
        + (\tau +1) F_{2,1}^V \left( t \right),
    \end{align}
\end{mysubeq} \par\noindent
where $\tau=-t/(4 M^2)$ and $G_{E0}$, $G_{E2}$, $G_{M1}$, and $G_{M3}$ represent the 
electric-monopole, electric-quadrupole, magnetic-dipole, and magnetic-octupole form 
factors, respectively. The intrinsic electromagnetic properties, including the electric 
charge, magnetic moment, electric-quadrupole moment, and magnetic-octupole moment, are 
obtained in the forward limit, $t=0$. The electric-monopole and magnetic-dipole form 
factors give the corresponding electric charge and magnetic radii of the particle as \cite{Leinweber:1992hy}
\begin{equation}\label{EMRadius}
    \left\langle r^2 \right\rangle_{E0} = \frac{6}{G_{E0}(0)} \frac{d}{dt}G_{E0}(t) \Big \vert _{t=0}, \quad 
    \left\langle r^2 \right\rangle_{M1} = \frac{6}{G_{M1}(0)} \frac{d}{dt}G_{M1}(t) \Big \vert _{t=0}.
\end{equation}

Similarly to EMFFs, the matrix element of the EMT current can be written as \cite{Kim:2020lrs}
\begin{equation}\label{EMT}
    \begin{split}
    &\left\langle p^\prime,\lambda^\prime \left| \hat{T}^{\mu \nu}_a(0)
    \right| p,\lambda\right\rangle  \\
    &~~~=-\bar{u}_{\alpha ^\prime}\left(p^\prime,\lambda^\prime\right) \bigg [\frac{P^\mu P^\nu}{M} \left(g^{\alpha'  \alpha} F_{1,0}^{T,a} (t)
    -\frac{q ^{\alpha' } q ^{\alpha}}{2 M^2}F_{1,1}^{T,a} (t) \right)
    + \frac{ \left({q }^\mu {q }^\nu- {g}^{\mu  \nu } q ^2\right) }{4M}
    \left({g}^{\alpha'  \alpha}F_{2,0}^{T,a} (t)-\frac{{q }^{\alpha '}
    {q }^{\alpha}}{2 M^2}F_{2,1}^{T,a} (t)\right)  \\
    &~~~~~+ M g^{\mu  \nu } \left(
    g^{\alpha'  \alpha}F_{3,0}^{T,a} (t)-\frac{ q^{\alpha'}
    q^{\alpha}}{2 M^2}F_{3,1}^{T,a} (t)\right)
    + \frac{i {P}^{ \{ \mu } \sigma ^{\nu \} \rho} q_\rho}{2M} \left(g^{\alpha'  \alpha}F_{4,0}^{T,a} (t)
    -\frac{q ^{\alpha' } q ^{\alpha}}{2 M^2}F_{4,1}^{T,a} (t)\right) \\
    &~~~~~- \frac{1}{M} \left({q }^{\{ \mu} g^{\nu \} \{ \alpha '} {q }^{\alpha \}}
    -2 q^{\alpha' } q^{\alpha} g^{\mu  \nu }
    - g^{\alpha ' \{ \mu } g^{\nu \} \alpha } q^2 \right) F_{5,0}^{T,a} (t)
    + M g^{\alpha ' \{ \mu } g^{\nu \} \alpha}F_{6,0}^{T,a} (t) \bigg ]
    u_\alpha \left( p,\lambda\right),
    \end{split}
\end{equation}
where $F^T_{i,j}=\underset{a}{\sum} F_{i,j}^{T,a}$ stand for the GFFs of the spin-3/2 
hadron and the conventions $a^{\{ \mu}b^{\nu \}} = a^\mu b^\nu + a^\nu b^\mu$ and 
$a^{ [ \mu}b^{\nu ] } = a^\mu b^\nu - a^\nu b^\mu$ are adopted. In the Breit frame, 
the gravitational multipole form factors (GMFFs) of the spin-3/2 particle can be 
expressed in terms of its GFFs $F^T_{i,j}$ as \cite{Kim:2020lrs}
\begin{mysubeq}\label{GFFs}
    \begin{align}
        {\varepsilon}_0 \left( t \right) & = F^T_{1,0}(t)  + \frac{t}{6 M^2}\biggl[
        - \frac{5}{2} F^T_{1,0}(t) - F^T_{1,1}(t) - \frac{3}{2} F^T_{2,0}(t) + 4 F^T_{5,0}(t) + 3 F^T_{4,0} \biggr]\notag\\
        &+ \frac{t^2}{12M^4} \biggl[ \frac{1}{2} F^T_{1,0}(t) + F^T_{1,1}(t) + \frac{1}{2} F^T_{2,0}(t)
        + \frac{1}{2} F^T_{2,1}(t) - 4 F^T_{5,0}(t) - F^T_{4,0}(t) - F^T_{4,1}(t) \biggr]\notag\\
        &+ \frac{t^3}{48M^6} \biggl[ - \frac{1}{2} F^T_{1,1}(t) - \frac{1}{2} F^T_{2,1}(t)
        + F^T_{4,1}(t) \biggr],\\
        {\varepsilon}_2(t) & = - \frac{1}{6} \biggl[ F^T_{1,0}(t) + F^T_{1,1}(t) -4 F^T_{5,0}(t)  \biggr]\notag\\
        & + \frac{t}{12M^2} \biggl[ \frac{1}{2} F^T_{1,0}(t) + F^T_{1,1}(t) + \frac{1}{2}F^T_{2,0}(t)
        + \frac{1}{2}F^T_{2,1}(t) - 4 F^T_{5,0}(t) -F^T_{4,0} -F^T_{4,1}(t)  \biggr]\notag\\
        & + \frac{t^2}{48M^4} \biggl[ -\frac{1}{2} F^T_{1,1}(t) - \frac{1}{2}F^T_{2,1}(t) + F^T_{4,1}(t) \biggr], \\
        \mathcal{J}_1(t) &= F^T_{4,0}(t)
        - \frac{t}{5M^2} \biggl[ F^T_{4,0}(t) + F^T_{4,1}(t) + 5F^T_{5,0}(t) \biggr]
        + \frac{t^2}{20M^4}F^T_{4,1}(t),\\
        \mathcal{J}_3(t) &= - \frac{1}{6} \biggl[ F^T_{4,0}(t) + F^T_{4,1}(t) \biggr] + \frac{t}{24M^2}F^T_{4,1}(t), \\
        D_{0}(t) &= F^T_{2,0}(t) - \frac{16}{3}F^T_{5,0}(t) -\frac{t}{6M^{2}}\bigg{[} F^T_{2,0}(t) +F^T_{2,1}(t) -4F^T_{5,0}(t)\bigg{]}
        +\frac{t^{2}}{24M^{4}}F^T_{2,1}(t),\\
        D_{2}(t) &=\frac{4}{3} F^T_{5,0}(t), \\
        D_{3}(t) &= \frac{1}{6}\bigg{[}-F^T_{2,0}(t) -F^T_{2,1}(t) + 4 F^T_{5,0}(t)\bigg{]}+\frac{t}{24M^{2}}F^T_{2,1}(t),
    \end{align}
\end{mysubeq}\par\noindent
where the non-conserving terms, $F^T_{3,0(1)}$ and $F^T_{6,0}$ are simply ignored 
because they should vanish if we add the gluon contributions explicitly.
In Eq.~\eqref{GFFs}, ${\varepsilon}_{0(2)}$ and $\mathcal{J}_{1(3)}$ stand for the 
energy-monopole (-quadrupole) and angular momentum-dipole (-octupole) 
form factors, respectively. $D_{0 (2,3)}$ are regarded as the form factors associated 
with the internal pressures and shear forces \cite{Polyakov:2018zvc}. Like the 
electromagnetic radii defined in Eq.~\eqref{EMRadius}, there is a 
corresponding mass radius
\begin{equation}\label{MassRadius}
    \langle r^2 \rangle_M = \frac{6}{\varepsilon_0(0)} \frac{d}{dt} \left. \varepsilon_0(t) \right|_{t=0}.
\end{equation}

Moreover, to get the densities in the coordinate space, one may calculate the Fourier 
transformations of GMFFs. The corresponding $00$- and $ij$- components of the static 
EMT are~\cite{Kim:2020lrs}
\begin{equation}\label{eqT00}
    T^{00}(\bm{r}, \lambda', \lambda) = \mathcal{E}_0(r) \delta_{\lambda' \lambda} + \mathcal{E}_2(r) \hat{Q}^{l m}_{\lambda' \lambda} Y^{l m}_2 (\Omega_r),
\end{equation}
\begin{equation}\label{eqTij}
        T^{ij}(\bm{r}, \lambda', \lambda) = p_0(r) \delta^{i j} \delta_{\lambda' \lambda} + s_0(r) Y^{i j}_2 \delta_{\lambda' \lambda},
\end{equation}
where $\hat{Q}^{l m}$ and $Y^{l m}_2 (\Omega_r)$ are the quadrupole spin operator and 2-rank irreducible tensor as defined in Ref.~\cite{Kim:2020lrs}, respectively.
Here we neglect the high order terms $p_{2,3}$ and $s_{2,3}$ in $T^{ij}$ for simplicity.
The energy-monopole and -quadrupole densities can be further expressed as \cite{Kim:2020lrs}
\begin{equation}\label{epsi}
    \mathcal{E}_0(r) = M \widetilde{{\varepsilon}}_0(r), \qquad
    \mathcal{E}_2(r) = - \frac{1}{M} r \frac{d}{dr}\frac{1}{r} \frac{d}{dr}
    \widetilde{{\varepsilon}}_2(r),
\end{equation}
with 
\begin{eqnarray}\label{Fourie}
\widetilde{{\varepsilon}}_{0,2}(r) = \int \frac{d^3 q}{(2\pi)^3} e^{-i \bm{q} 
\cdot \bm{r}} {\varepsilon}_{0,2}(t),
\end{eqnarray}
being the densities in coordinate $r$-space.
Ref.~\cite{Polyakov:2018zvc} argued that the static $T^{i j}(\bm{r})$ may involve 
the pressure and shear force information in contrast to the classical mechanics 
for the continuous media. Then
\begin{equation}\label{force}
  \begin{split}
    p_n (r) & = \frac{1}{6 M} \frac{1}{r^2} \frac{d}{d r} r^2 \frac{d}{d r}
    \tilde{D}_n (r),\\
    s_n (r) & = - \frac{1}{4 M} r \frac{d}{d r} \frac{1}{r} \frac{d}{d r}
    \tilde{D}_n (r),
  \end{split}
\end{equation}
where
\begin{equation}
    \begin{split}
        \tilde{D}_0 (r) & = \int \frac{d^3 q}{(2 \pi)^3} e^{- i \bm{q} \cdot \bm{r}} D_0 (t),\\
        \tilde{D}_2 (r) & = \int \frac{d^3 q}{(2 \pi)^3} e^{- i \bm{q} \cdot \bm{r}} D_2 (t) + \frac{1}{M^2}\left( \frac{d}{dr}\frac{d}{dr}- \frac{2}{r}\frac{d}{dr} \right)\int \frac{d^3 q}{(2 \pi)^3} e^{- i \bm{q} \cdot \bm{r}} D_3 (t),\\
        \tilde{D}_3 (r) & = -\frac{2}{M^2} \left( \frac{d}{dr}\frac{d}{dr}- \frac{3}{r}\frac{d}{dr} \right) \int \frac{d^3 q}{(2 \pi)^3} e^{- i \bm{q} \cdot \bm{r}} D_3 (t).
    \end{split}
\end{equation}
Moreover, there is an equilibrium relation between the pressure and shear force densities
\begin{equation}\label{equilibrium1}
    \frac{2}{3} \frac{d s_n(r)}{dr} +2 \frac{s_n(r)}{r}+\frac{d p_n(r)}{dr}=0, \quad \text{with} \quad n=0,2,3.
\end{equation}

Another interest is the angular momentum density, which is obtained from the $0k$-components of the static EMT as \cite{Kim:2020lrs}
\begin{equation}\label{rhor}
    \rho_J(r)= - \frac{1}{3} r \frac{d}{d r} \int \frac{d^3 q}{(2 \pi)^3} e^{- i \bm{q \cdot r}} \mathcal{J}_1(t),
\end{equation}
which describes the angular momentum distribution in  coordinate space and gives 
the total spin by the integral in the 3D space.

\subsection{Quark-diquark approach}

\quad\quad We know that the $\Omega^-$ hyperon, which has the quantum number of 
$I(J^P) =0 (3/2^+)$, is composed of three $s$ quarks. It is convenient to consider 
$\Omega^-$ as a bound state with one $s$ quark and one diquark. The latter consists 
of two $s$ quarks and has $J^P =1^+$. We explicitly consider the internal structure 
of the axial-vector diquark in order to give a more precise description. This 
approach is consistent with other relativistic and covariant quark-diquark 
approaches \cite{Meyer:1994cn,Keiner:1995bu} and was employed in our previous 
work \cite{Fu:2022rkn}.

\begin{figure}[htbp]
\centering
    \subfigure[]{\includegraphics[scale=0.5]{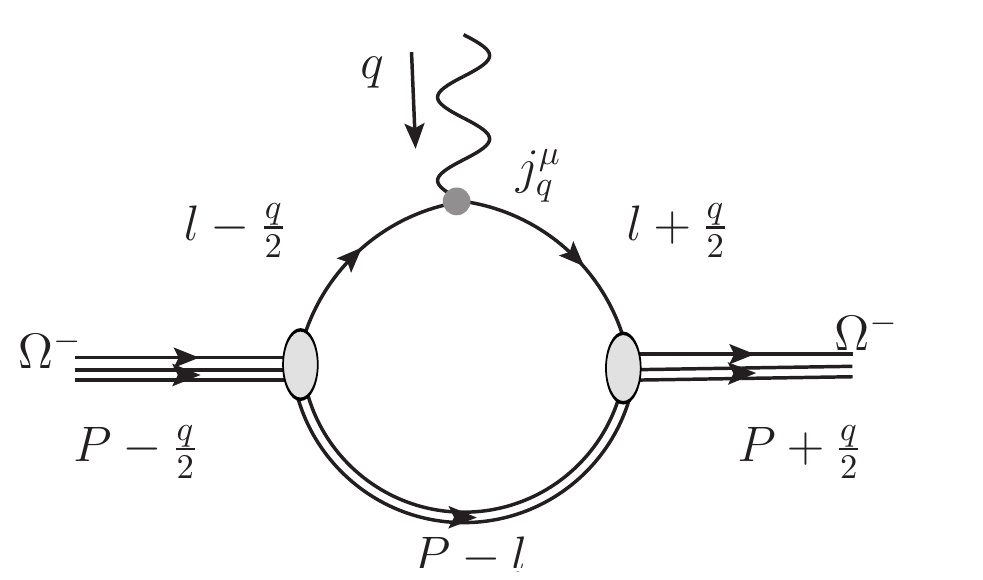}}
    \subfigure[]{\includegraphics[scale=0.5]{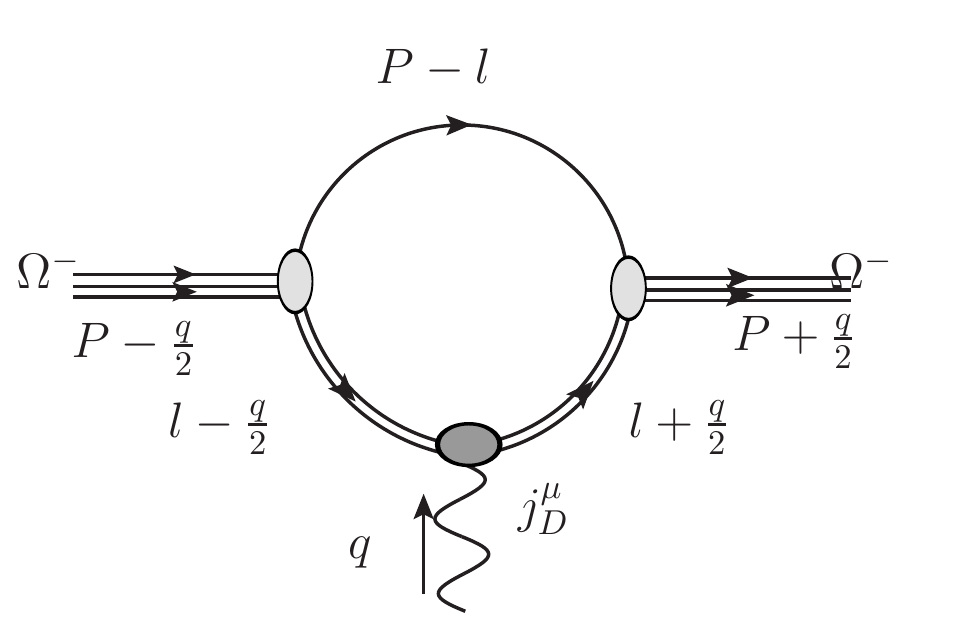}}
    \subfigure[]{\includegraphics[scale=0.5]{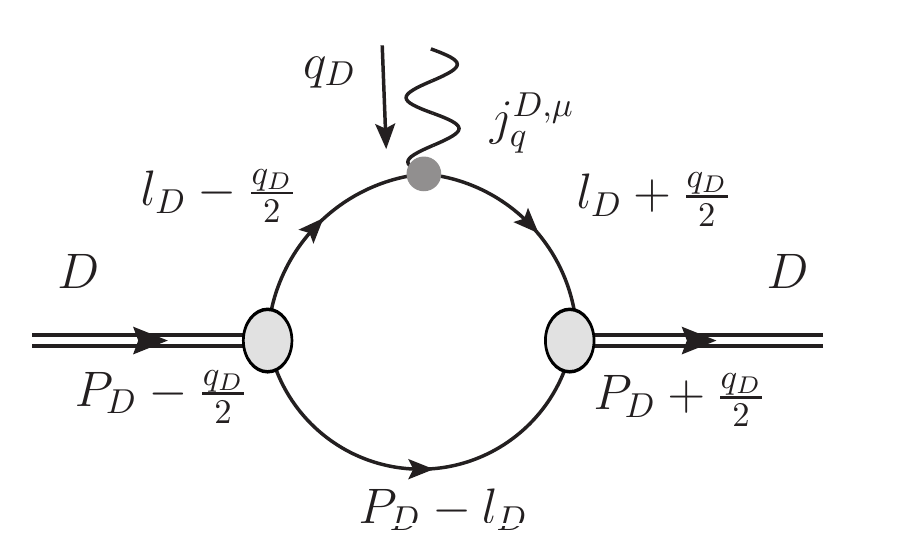}}
    \caption{\small{Feynman diagrams for the electromagnetic current of $\Omega^-$, (a) and (b), and of the diquark (c). The left and middle panels stand for the contributions of quark (single line) and diquark (double line) to $\Omega^-$, respectively.}}
    \label{feynman}
\end{figure}

Here we briefly show our calculation of the EMFFs for $\Omega^-$ in the 
quark-diquark approach. EMFFs can be obtained from the matrix element of the 
electromagnetic current attached to $\Omega^-$. This process is displayed in 
Figs.~\ref{feynman} (a) and (b).
Thus, the matrix element is expressed as the sum of the quark and diquark contributions as
\begin{equation}
    \left\langle p^\prime,\lambda^\prime \left| \hat{J}^{\mu}(0) \right| p,\lambda\right\rangle = \left\langle p^\prime,\lambda^\prime
    \left| \hat{J}^{\mu}_{q}(0) \right| p,\lambda\right\rangle + \left\langle p^\prime,\lambda^\prime \left| \hat{J}^{\mu}_D(0)
    \right| p,\lambda\right\rangle.
\end{equation}

One can get the quark contribution from the Feynman diagram \ref{feynman} (a) as
\begin{equation}\label{Omegaquarkj}
    \begin{split}
    \left\langle p^\prime,\lambda^\prime \left| \hat{J}^{\mu}_{q}(0)
    \right| p,\lambda\right\rangle = & - Q^e_{q} e \bar{u}_{\alpha'}(p',\lambda') {\left( -i c^2 \right)} \\
    ~~~&\times \int \frac{d^4 l}{(2 \pi)^4}\frac{1}{\mathfrak{D}}
    \Gamma^{\alpha' \beta'} \left( \slashed{l}+\frac{\slashed{q}}{2}+m_q \right)
    g_{\beta' \beta} \gamma^\mu \left( \slashed{l}-\frac{\slashed{q}}{2}
    +m_q \right)\Gamma^{\alpha \beta} u_{\alpha}(p,\lambda),
    \end{split}
\end{equation}
where $Q_q^e$ is the electric charge number carried by the quark participating in 
the interaction and
\begin{equation}\label{OmegaD}
    \begin{split}
    \mathfrak{D}=& [\left( l-P \right) ^2 -m_R^2+i \epsilon]^2
    [\left(l-P\right)^2-m_D^2+i \epsilon]\\
    & \times \biggl[\left( l-\frac{q}{2} \right) ^2 -m_R^2+i \epsilon\biggr]
    \biggl[\left( l+\frac{q}{2} \right) ^2 -m_R^2+i \epsilon\biggr]\biggl[\left(l+ \frac{q}{2} \right)^2 - m_q^2+i \epsilon\biggr]
    \biggl[\left(l- \frac{q}{2} \right)^2 - m_q^2+i \epsilon\biggr].
    \end{split}
\end{equation}
In Eq.~\eqref{Omegaquarkj}, the effective vertex is employed as
\begin{equation}\label{vertexfunction}
    \Gamma^{\alpha \beta} = g^{\alpha \beta} + c_2 \gamma^\beta \Lambda^\alpha
    + c_3 \Lambda^\beta \Lambda^\alpha,
\end{equation}
where $\Lambda$ is the relative momentum between the quark and diquark, and the 
superscript $\alpha$ ($\beta$) represents the index of the $\Omega^-$ (diquark). 
$m_q$ and $m_D$ are the masses of the quark and the diquark, respectively.
The couplings, $c_2$ and $c_3$ can be determined by fitting to the LQCD results of 
EMFFs. To avoid the loop integral divergence, we employ one 
simple regularization at each vertex, i.e.  we add a scalar function
\begin{equation}\label{vertexfunction2}
    \Xi (p_1,p_2)=\frac{c}{[ p_1^2 -m_R^2+i \epsilon][ p_2^2 -m_R^2+i \epsilon]},
\end{equation}
where $m_R$ is a cutoff mass parameter. In Eq.~\eqref{vertexfunction2} the 
parameter $c$ is fixed in order to give the electric charge number of 
$\Omega^-$ at $t=0$. It should be mentioned that this simplification may 
break the gauge invariant slightly, however it is simpler than other 
sophisticated methods, such as the Pauli-Villars  regularization 
\cite{Pauli:1949zm}.

According to Fig.~\ref{feynman} (b), the diquark contribution can be expressed 
as
\begin{equation}\label{OmegaDiquarkj}
    \left\langle p^\prime,\lambda^\prime \left| \hat{J}^{\mu}_D(0)
    \right| p,\lambda\right\rangle
    = - Q^e_{D} e \bar{u}_{\alpha'}(p',\lambda') {i c^2} \int \frac{d^4 l}
    {(2 \pi)^4}\frac{1}{\mathfrak{D}'}
    \Gamma^{\alpha'}_{~\beta'} \left( \slashed{P}-\slashed{l}+m_q \right)
     j_D^{\mu,\beta' \beta} \Gamma^{~\alpha}_{\beta} u_{\alpha}(p,\lambda),
\end{equation}
where $Q_D^e$ is the electric charge number carried by the diquark. The diquark 
electromagnetic current then can be calculated explicitly 
from Fig.~\ref{feynman} (c) 
as
\begin{equation}\label{Diquarkj}
    \sum_{q} \left\langle p_D^\prime,\lambda_D^\prime \left| \hat{J}^{\mu}_{q}(0) 
    \right| p_D,\lambda_D\right\rangle = 
    - \epsilon^*_{\beta ^\prime}\left(p_D^\prime,\lambda_D^\prime\right) 
    j_D^{\mu, \beta' \beta} \epsilon_\beta \left( p_D,\lambda_D \right),
\end{equation}
where $\epsilon_\beta \left( p_D,\lambda_D \right)$ is the spin-1 diquark 
field and we simply assume that the axival-vector diquark is on-shell. 
The calculation details of Eq.~\eqref{Diquarkj} are referred to 
Ref.~\cite{Fu:2022rkn}. 

Finally, the calculation of the GFFs of the $\Omega^-$ hyperon is similar to 
that of EMFFs replacing the electromagnetic current $j^\mu$ by the EMT 
current $T^{\mu \nu}$~\cite{Fu:2022rkn}.

\section{Numerical results}\label{sectionresults}
\subsection{Determination of parameters}

\quad\quad  We know that the formal FFs should be extracted from the integral 
in Eqs.~\eqref{Omegaquarkj} and \eqref{OmegaDiquarkj} by using the on-shell 
identities of the Rarita-Schwinger fields \cite{Cotogno:2019vjb,Fu:2022rkn}. 
Moreover, we also need to input the $\Omega^-$ mass $M$, $s$ quark mass $m_q$, and 
diquark mass $m_D$ as the model parameters. To ensure that $\Omega^-$ and the 
diquark are bound states, $M$, $m_q$, and $m_D$ need to satisfy the relation, 
$M< m_q+m_D < 3 m_q$. Here, we choose $M=1.672 ~\text{GeV}$ 
\cite{ParticleDataGroup:2022pth}, $m_q = 0.6 ~\text{GeV}$, 
and $m_D=1.15 ~\text{GeV}$. In addition, other model parameters, the cutoff 
mass $m_R$ and the couplings $c_{2(3)}$ in Eqs.(\ref{OmegaD}-\ref{vertexfunction2}), 
can be modulated to obtain more reasonable form factors, comparing to those of the 
LQCD calculations. Thus, we finally choose $m_R =2.2 \,\text{GeV} \gtrsim M$, 
$c_2=0.306 \,\text{GeV}^{-1}$, and $c_3=0.056 \text{GeV}^{-2}$. These three 
parameters and the input masses are listed in Table~\ref{parameterstable}.
\begin{table}[h]
    \centering
	\begin{tabular}{  c  c  c  c  c  c  c }
		\hline
		\hline
		\specialrule{0em}{0pt}{2pt}
		$M/\text{GeV}$ & $m_q/\text{GeV}$ & $m_D/\text{GeV}$ & $m_R/\text{GeV}$	& $c_2/\text{GeV}^{-1}$	& $c_3/\text{GeV}^{-2}$\\
		\hline
		1.672   & 0.6   & 1.15  & 2.2	& 0.306	& 0.056	\\
		\hline
		\hline
	\end{tabular}
    \caption{\small{The parameters used in this work.}}
    \label{parameterstable}
\end{table}

\begin{figure}[h]
\centering
    \includegraphics[height=4.5cm]{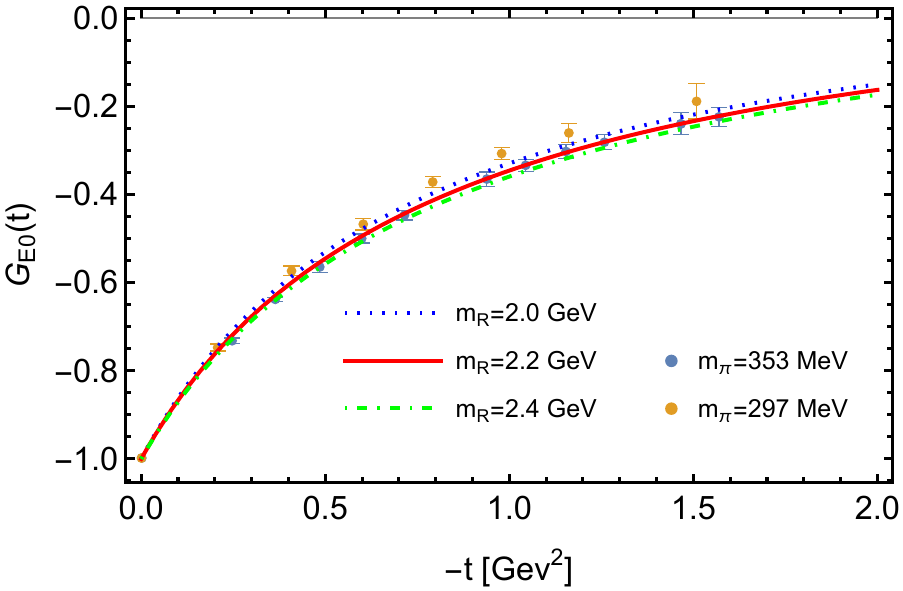}
    \caption{\small{The comparison of $G_{E0}$ with LQCD to different $m_R$ when $c_2=0.306 \,\text{GeV}^{-1}$ and $c_3=0.056 \text{GeV}^{-2}$.}}
    \label{FiguremR}
\end{figure}

Figure~\ref{FiguremR} gives the comparison of our electric form factor $G_{E0}$ to 
the results of LQCD~\cite{Alexandrou:2010jv} with different $m_R$. We conclude that 
the results are not sensitive to the parameter $m_R$.
Furthermore, we find that the parameters $c_{2(3)}$ make a significant impact on the 
high-order multipoles form factors, such as the electric-quadrupole, 
magnetic-octupole, energy-quadrupole, and angular momentum-octupole form factors 
as discussed in Ref.~\cite{Fu:2022rkn}, especially on even higher-order 
multipole magnetic and angular momentum -octupole form factors.

\subsection{Results of EMFFs of the $\Omega^-$ hyperon}

\begin{figure}[h]
\centering
    \includegraphics[height=4.5cm]{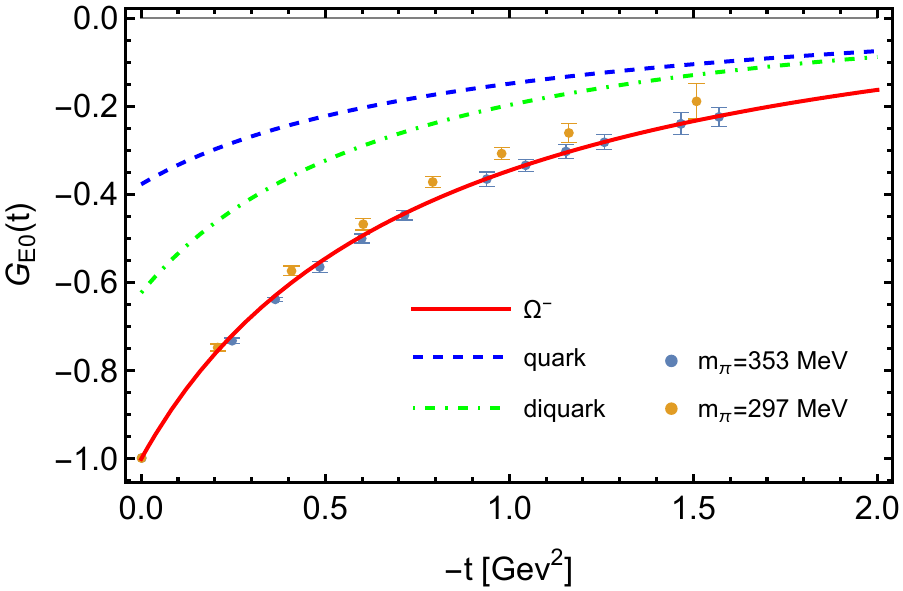} \quad
    \includegraphics[height=4.45cm]{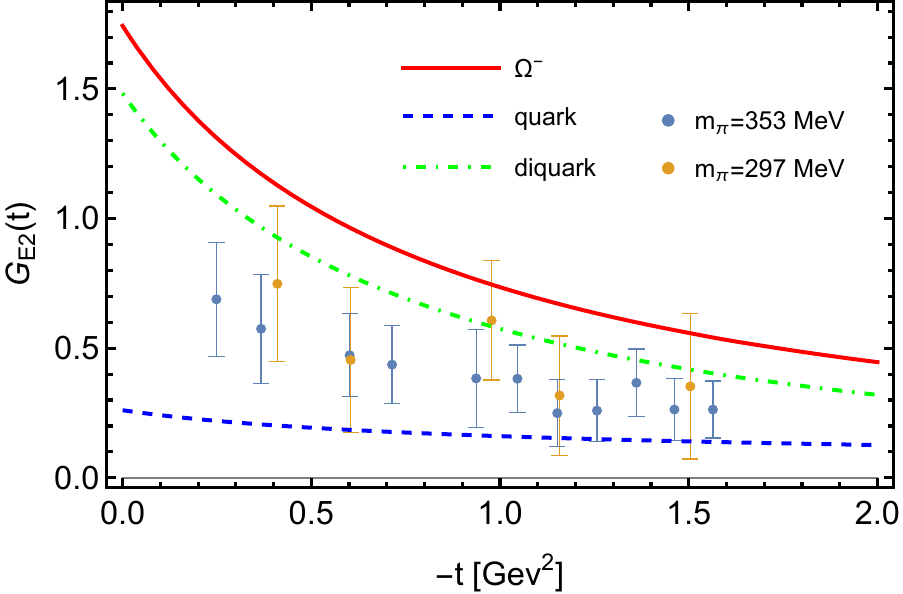}\\
    \includegraphics[height=4.5cm]{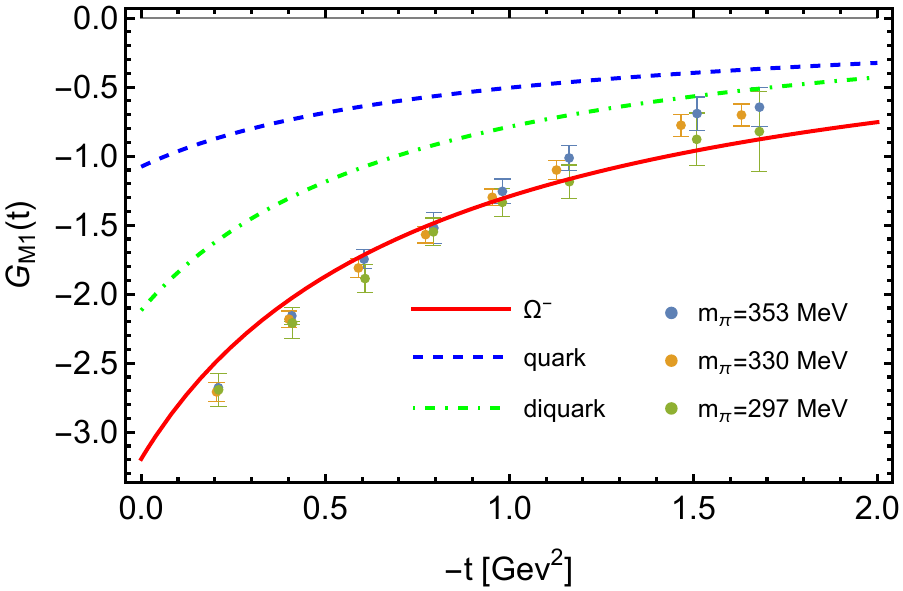} \quad
    \includegraphics[height=4.45cm]{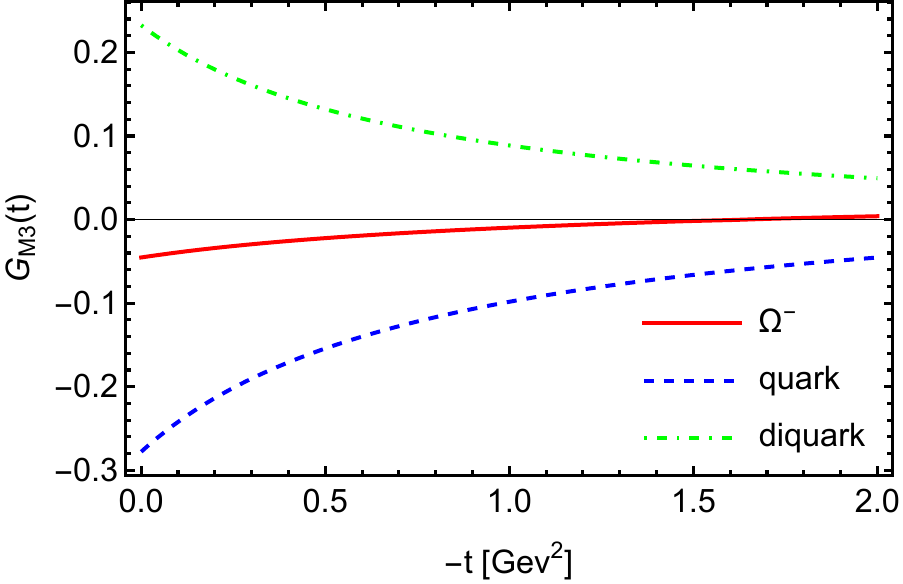}
    \caption{\small{Our EMFFs in comparison with the results of  LQCD~\cite{Alexandrou:2010jv}. The dashed, dotted-dashed and solid lines represent the quark, diquark, and total contributions, respectively.}}
    \label{FigureEMFFs}
\end{figure}

\quad\quad Once the parameters are determined, the EMFFs of $\Omega^-$, including 
the electric-monopole, magnetic-dipole, electric-quadrupole and magnetic-octupole 
form factors, can be calculated. Our results are compared with the LQCD 
calculations~\cite{Alexandrou:2010jv} in Fig.~\ref{FigureEMFFs}. In the 
Fig.~\ref{FigureEMFFs}, the contributions from the quark and diquark are 
explicitly displayed. We find that both our calculation and LQCD result are 
consistent with each other. In particular, our electric-monopole and 
magnetic-dipole form factors match the LQCD results better. Since the electric 
charge carried by the diquark is twice that of the quark, the ratio between the 
diquark and quark contributions is about $2$, as $-t$ tends to $0$. In the forward 
limit, Fig.~\ref{FigureEMFFs} gives the magnetic moment $\mu_{\Omega^-}=G_{M1} \frac{M_N}{M} \mu_N$, electric-quadrupole moment 
$\mathcal{Q}_{\Omega^-}= G_{E2}(0) \frac{\lvert e \rvert}{M^2}$, 
and magnetic-octupole moment $\mathcal{O}_{\Omega^-}=G_{M3}(0) 
(\frac{M_N}{M})^3 \mathcal{O}_{N}$. The physical quantities of 
$\mu_N$ and $\mathcal{O}_N$ with the subscript $N$ stand for the corresponding 
nuclear properties and $M_N$ being the proton mass.
A comparison of our results with those of the different models is also shown in 
Tab.~\ref{TableEMFFs}.
From Tab.~\ref{TableEMFFs}, we see that our magnetic moment $\mu_{\Omega^-} = -1.8 \, 
\mu_N$ is slightly less than the experiment value $-2.02 (5) \, \mu_N$ and is 
close to the LQCD and $\chi$QSM results. Moreover, our electric-quadrupole moment
is of the same order as others. We know that the electric-quadrupole form factors 
show the 3D electric charge distribution shape of the system, and 
$\mathcal{Q}_{\Omega^-}>0$ implies that the electric charge distribution of 
$\Omega^-$ is a prolate ellipsoid.  In addition, the electromagnetic radii from 
~\eqref{EMRadius} are important quantities for us to apprehend the electromagnetic 
properties of the system, and they are
\begin{equation}
    \langle r^2 \rangle_{E0} = 0.352 \, \text{fm}^2 \quad \text{and} \quad \langle r^2 \rangle_{M1} = 0.322 \, \text{fm}^2,
\end{equation}
for the $\Omega^-$ hyperon. Our results are comparable with other model calculations 
as shown in the Tab.~\ref{TableEMFFs}. One can conclude that the magnetic radius 
is smaller than the electric charge radius and this relation is also in agreement 
with other model predictions except for the RQM calculation \cite{Ramalho:2009gk}.

\begin{table}[h]
    \centering
	\begin{tabular}{ c  c  c  c  c  c }
		\hline
		\hline
		\specialrule{0em}{0pt}{2pt}
		\text{  }	& $\mu_{\Omega^-}/\mu_{N}$	& $\mathcal{Q}_{\Omega^-}/\text{fm}^2$    & $\mathcal{O}_{\Omega^-}/\mathcal{O}_N$   & $\langle r^2\rangle_{E0}/\text{fm}^2$  & $\langle r^2 \rangle_{M1}/\text{fm}^2$\\
		\specialrule{0em}{1pt}{0pt}
		\hline
		\specialrule{0em}{0pt}{1pt}
		\text{this work}	& -1.8	& 0.024  &-0.008	& 0.352  &0.322	\\
		\hline
		PDG~\cite{ParticleDataGroup:2022pth}   & -2.02(5)   & -   &-  & -  & - \\
		LQCD~\cite{Leinweber:1992hy}    &-1.73(22)   &0.0042(56)  & $-9.989\pm 2.65$  & 0.226(16)  & 0.226(16)\\
		LQCD~\cite{Alexandrou:2010jv}   & -1.835(94)    & 0.0133(57) &-   & 0.355(14)   & 0.286(31)\\
		LQCD~\cite{Boinepalli:2009sq}   &-1.697(65) & 0.0086(12)    &0.2(1.2)  & 0.307(15)   & -\\
		$\chi$PT~\cite{Butler:1993ej}   &-1.94(22)  & 0.009(5)  &-  &-  &-\\
		$\chi$PT~\cite{Li:2016ezv}  &-2.02(5)   & -     & -     & 0.70(12)    & - \\
		$1/N_c$~\cite{Luty:1994ub,Buchmann:2018nmu}  & -1.94     & 0.018     & -0.65     & -     & - \\
		RQM~\cite{Ramalho:2009gk}  & -2.02(5)     & -     & -     & 0.22     & 0.27 \\
		QCDSR~\cite{Lee:1997jk}  & -1.49(45)     & -     & -     & -     & - \\
        QCDSR~\cite{Aliev:2009pd}  & -     & 0.12(4)     & 1.73(43)     & -     & - \\
		NRQM~\cite{Krivoruchenko:1991pm}  & -     & 0.028     & -     & -     & - \\
		$\chi$QM~\cite{Wagner:2000ii}  & -2.13     & 0.026     & -     & 0.61     & 0.53 \\
		$\chi$QSM~\cite{Kim:2019gka,Jun:2021bwx}  & -1.82     & 0.054     & -     & 0.832     & 0.582 \\
		\specialrule{0em}{1pt}{0pt}
		\hline
		\hline
	\end{tabular}
    \caption{\small{The magnetic moment, electric-quadrupole moment, magnetic-octupole moment, electric charge radius and magnetic radius in comparison with those from PDG~\cite{ParticleDataGroup:2022pth}, LQCD~\cite{Leinweber:1992hy,Alexandrou:2010jv,Boinepalli:2009sq}, $\chi$PT~\cite{Butler:1993ej,Li:2016ezv}, $1/N_c$ expansion~\cite{Luty:1994ub,Buchmann:2018nmu}, relativistic quark model~\cite{Ramalho:2009gk}, non-relativistic quark model~\cite{Krivoruchenko:1991pm}, QCD sum rules~\cite{Lee:1997jk,Aliev:2009pd}, $\chi$ quark model~\cite{Wagner:2000ii}, and chiral quark-soliton model~\cite{Kim:2019gka,Jun:2021bwx}.}}
    \label{TableEMFFs}
\end{table}

\subsection{Results of GMFFs of $\Omega^-$ baryon}

\begin{figure}[h]
\centering
    \includegraphics[height=3.5cm]{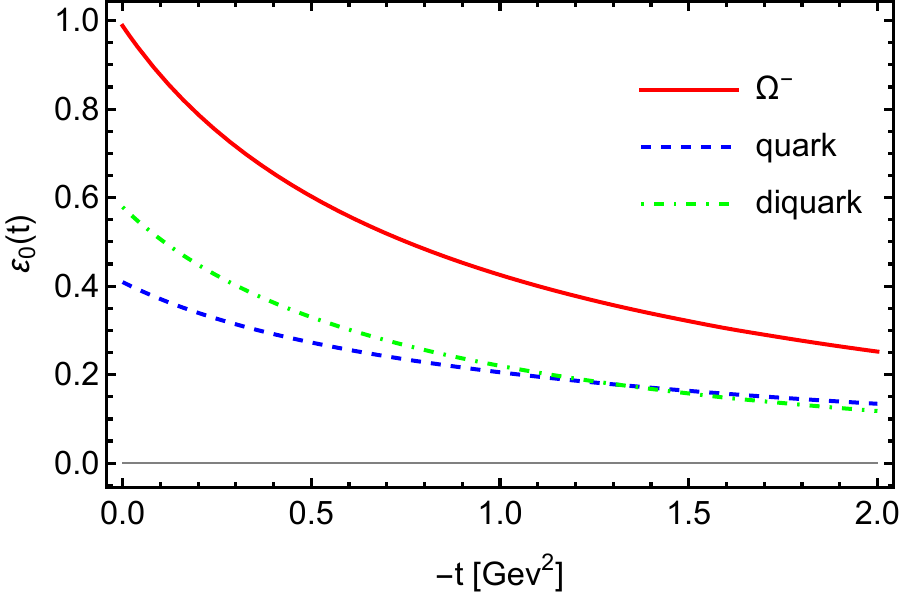}
    \includegraphics[height=3.5cm]{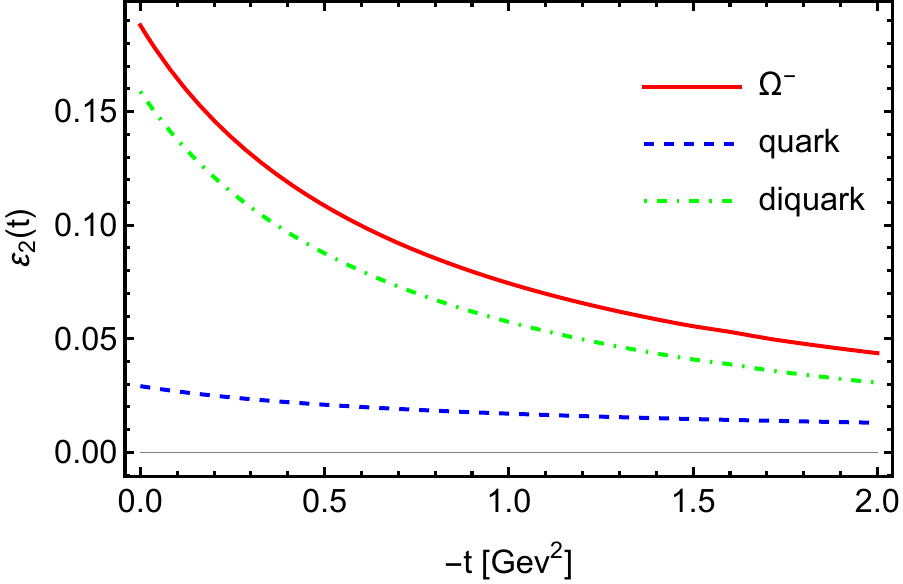}\\
    \includegraphics[height=3.5cm]{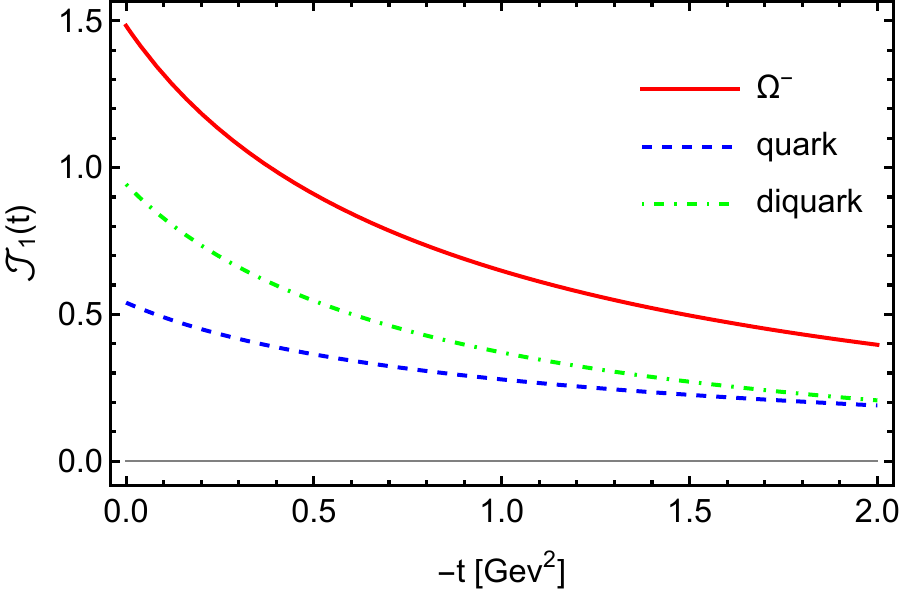}
    \includegraphics[height=3.5cm]{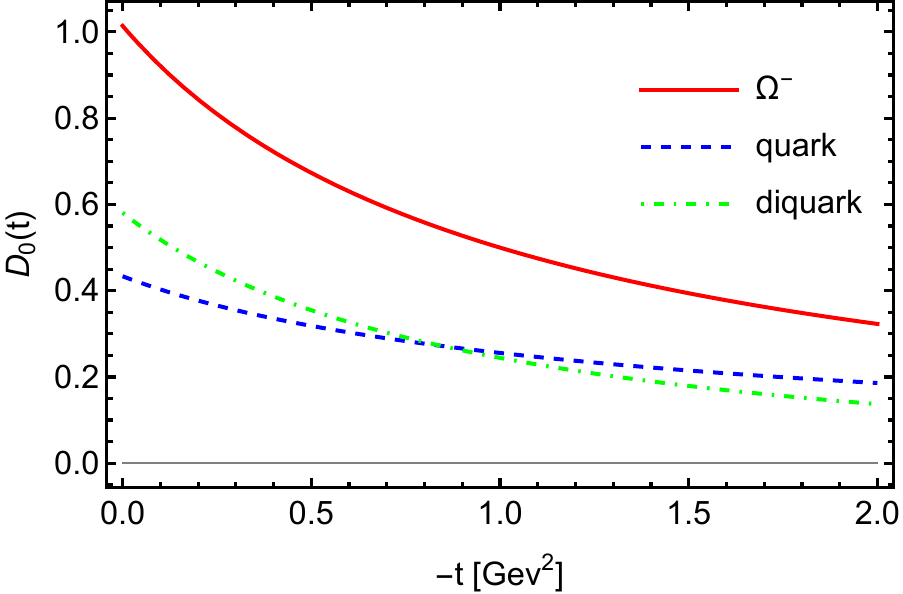}
    \caption{\small{The low order terms of the gravitational form factors of $\Omega^-$ as the functions of the squared momentum transfer $t$.}}
    \label{FigureGMFFs}
\end{figure}

\quad\quad Analogously, the matrix element of energy-momentum tensor gives GMFFs, 
which are expressed in terms of GFFs using the linear components in the Breit 
frame. By employing the same parameters and the same normalization, the GMFFs, 
including the energy-monopole $\varepsilon_0$, angular momentum-dipole 
$\mathcal{J}_1$, energy-quadrupole $\varepsilon_2$, angular momentum-octupole 
$\mathcal{J}_3$ form factors, and some other form factors such as $D_0$, $D_2$ and 
$D_3$, which may relate to the pressures and shear forces, can be obtained and 
the their low-order multipole terms are shown in Fig.~\ref{FigureGMFFs}.
In the forward limit $t=0$, the intrinsic mechanical properties of the $\Omega^-$ 
hyperon, like its mass, $\varepsilon_0(0)=0.988 \sim 1$, and spin, 
$\mathcal{J}_1(0)=1.483 \sim 3/2$, can be obtained in this approach. 
It is clearly seen that our obtained mass and spin are not the same as the exactly 
global physical quantities because the momentum-dependence regularization in 
Eq.~\eqref{vertexfunction2} violates the gauge invariance slightly. Similar to 
EMFFs, we find that the ratios between the diquark and quark contributions to 
the energy-monopole and to the angular momentum-dipole form factors are close to 2, 
especially for the small $-t$. This is intuitive because the mass and spin of the 
quark are practically about half of the diquark. It should be stressed that the 
shape of the energy distribution is another important property, thereupon we can 
conclude that the $\Omega^-$ is a prolate ellipsoid because of the positive 
$\varepsilon_2(0)$.  Finally, we get the mass radius of the $\Omega^-$ hyperon as
\begin{equation}
    \langle r^2 \rangle_M = 0.297 \, \text{fm}^2
\end{equation}
from Eq.~\eqref{MassRadius}. It is found that this mass radius is slightly smaller 
than the electromagnetic radii, $\langle r^2 \rangle_{E0}$ and $\langle r^2 
\rangle_{M1}$,  like our calculation for the $\Delta$ resonances \cite{Fu:2022rkn}. 

Finally, the $D$-term is also an essential mechanical quantity, which is defined as  
$D=D_0(0)$ and is argued to be negative and closely related to the stability of the 
system~\cite{Perevalova:2016dln}. Here, we get $D\sim 1.01$. It is positive and 
similar to the value for the $\Delta$ resonance in our previous 
calculation~\cite{Fu:2022rkn}. The possible interpretation of the positive $D$-term will be discussed in the following subsection.

\subsection{GMFFs in $r$-space}
\quad \quad
The local density distributions, including the energy densities \eqref{epsi}, 
angular momentum density \eqref{rhor}, and the internal forces \eqref{force}, 
can be obtained from the Fourier transformed form factors. To consider local 
particles, we simply employ a wave packet to describe the $\Omega^-$ hyperon.   
It should be addressed that Ref.~\cite{Epelbaum:2022fjc} concludes that the local 
density distributions must depend on the wave packet. Here, we simply employ an 
additional Gaussian-like wave packet $e^{\frac{t}{\lambda^2}}$ to describe the 
system \cite{Ishikawa:2017iym} as an approximation in Eqs.(10,12,14). 
In addition, this description can also guarantee the good convergence in the 
Fourier transformations. This additional wave packet may affect 
the radius definition \cite{Epelbaum:2022fjc,Alharazin:2022xvp},  however, 
this issue is not a priority in this work.

\begin{figure}[htbp]
\centering
    \includegraphics[height=4.5cm]{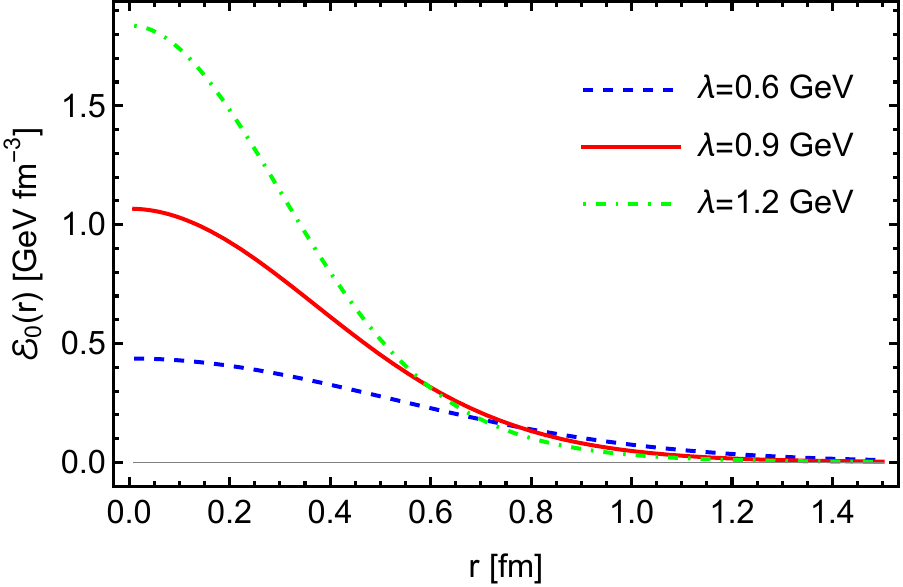}
    \includegraphics[height=4.5cm]{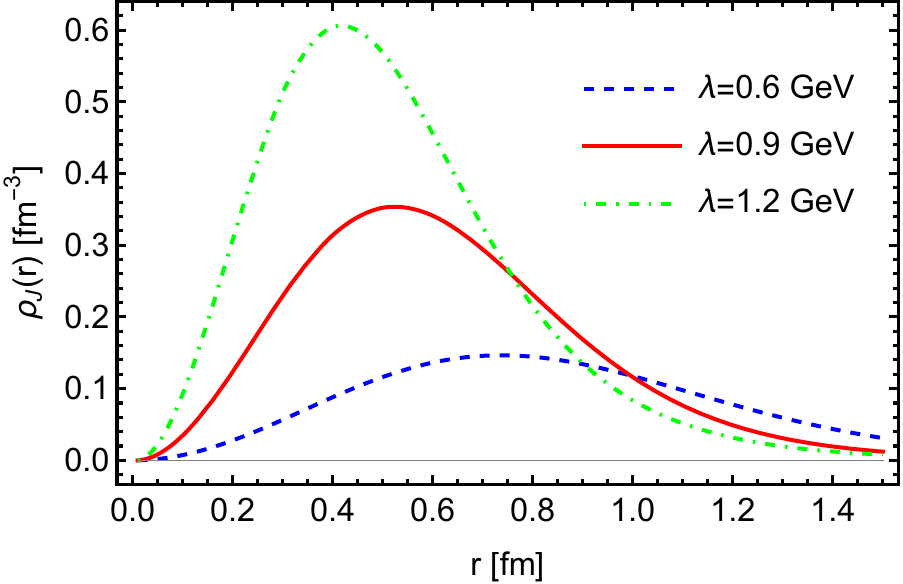}
    \caption{\small{The calculated energy-monopole (left panel) and angular momentum (right) 
    densities of $\Omega^-$ as the functions of $r$ with different $\lambda$.}}
    \label{FigureMassRRhoJ}
\end{figure}

It should be mentioned that the parameter $1/\lambda$ here characterizes the size of 
$\Omega^-$ and $\lambda$ has the mass dimension. Thus, one can conclude that 
the large $\lambda$ represents the small radius (the small $\lambda$ is opposite) 
according to the uncertainty principle. Thereupon the large $\lambda$ concentrates 
the densities close to the center (small $r$ region) and the small $\lambda$ to the 
contrary as shown in Fig.~\ref{FigureMassRRhoJ},
where we choose the reasonable $\lambda$ range $0.6 \,\text{GeV} < \lambda < 1.2 
\,\text{GeV}$. Furthermore, there are certainly some invariants in the energy and 
angular momentum densities in Fig.~\ref{FigureMassRRhoJ}. For example, 
the integrated result of $\rho_J(r)$ in the 3D coordinate space corresponds to the 
total spin and is independent of $\lambda$, and the integrated result of 
$\varepsilon_0(r)$ gives the mass term. Note that $\lambda = 0.9 \,\text{GeV}$ is 
employed in the following discussion.is not a priority

\begin{figure}[htbp]
\centering
    \includegraphics[height=8cm]{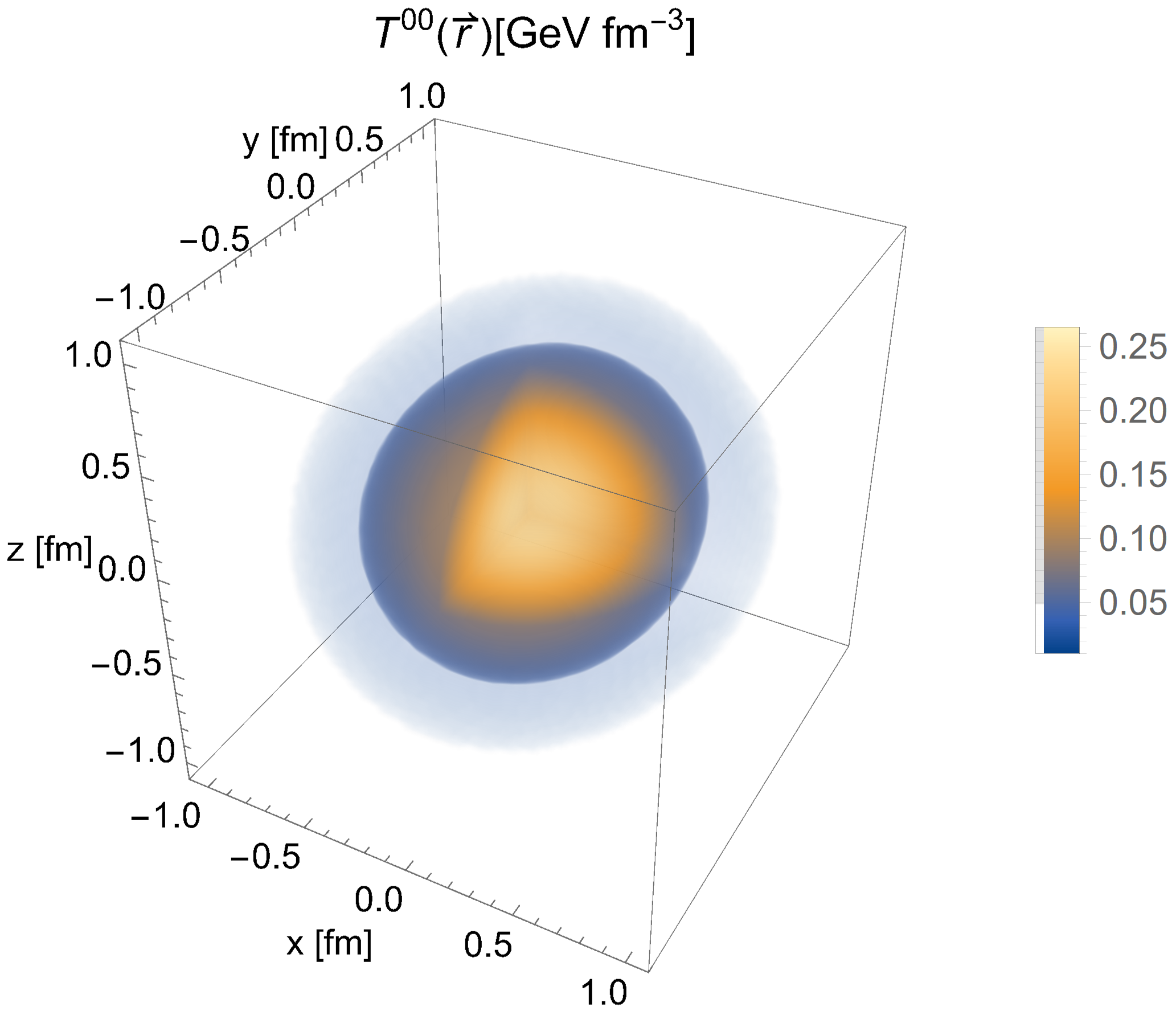}
    \caption{\small{The energy density using the polarization average.}}
    \label{FigureT00}
\end{figure}

The energy density in 3D space, from Eq.~\eqref{eqT00} taking the polarization 
average, is shown in Fig.~\ref{FigureT00}. One sees that the energy distribution 
has a prolate shape due to the positive energy-quadrupole form factor mentioned 
above.

\begin{figure}[htbp]
\centering
    \includegraphics[height=4.5cm]{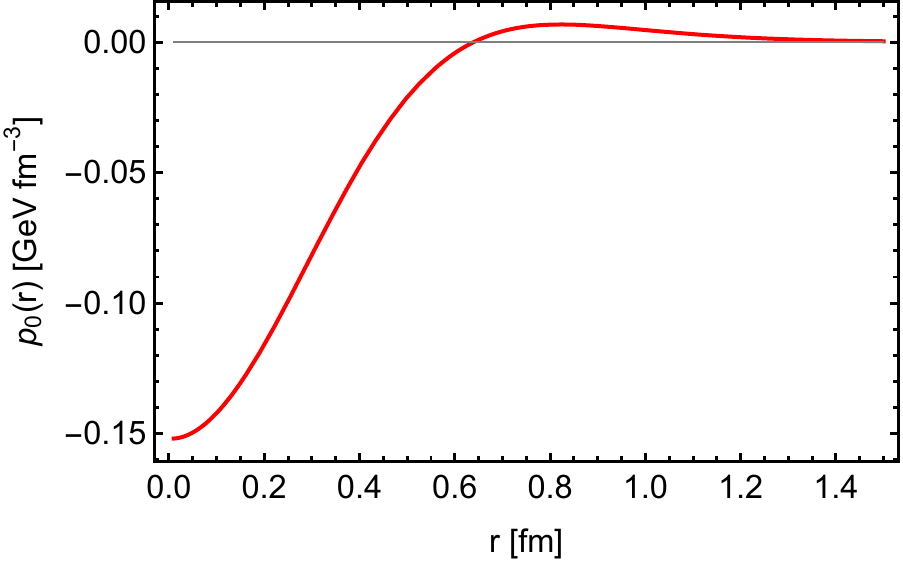}
    \includegraphics[height=4.5cm]{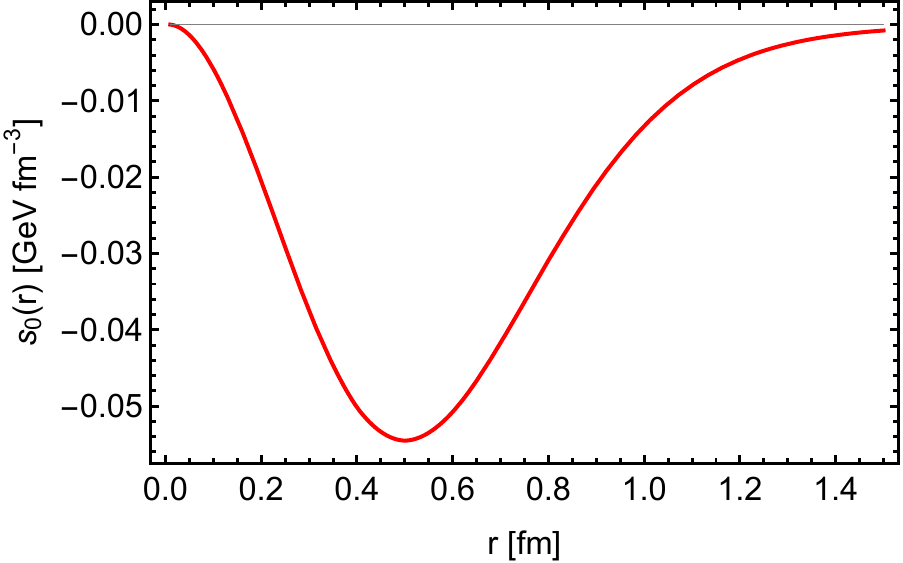}
    \caption{\small{The pressure (left panel) and shear force (right panel) of $\Omega^-$ as functions of $r$ when $\lambda=0.9 \,\text{GeV}$.}}
    \label{FigureForce0}
\end{figure}

The next relevant part is the $ij$-component of the static EMT, which describes the 
so-called pressures $p_n(r)$ and shear forces $s_n(r)$ argued in 
Refs.~\cite{Polyakov:2018zvc,Perevalova:2016dln} and is related to the GFFs.
According to Eq.~\eqref{force}, the pressure and shear force are shown in 
Fig.~\ref{FigureForce0} and they satisfy the equilibrium relation 
\eqref{equilibrium1}. In the left panel of Fig.~\ref{FigureForce0}, there is a 
crossing at about $r \sim 0.6 \,\text{fm}$, which is slightly larger than the 
mass radius and depends on $\lambda$. We conclude that that $r \sim 0.6 \,\text{fm}$ 
represents that there is a change in the pressure direction at the particle 
boundary.

\begin{figure}[htbp]
\centering
    \includegraphics[height=4.5cm]{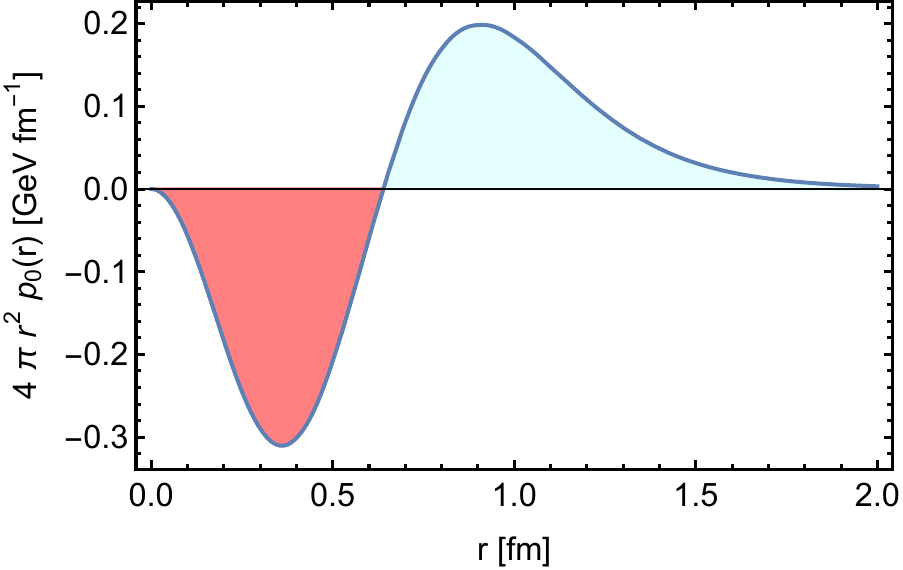}
    \caption{\small{The physical quantity $4 \pi r^2 p_0(r)$ as a function of $r$.}}
    \label{Figurep0r}
\end{figure}

Ref.~\cite{Perevalova:2016dln} stressed that the $D$-term is a fundamental and 
unknown quantity. It represents the stability of the system. The $D$-term 
is negativity since the corresponding inner force must be outward~\cite{Perevalova:2016dln}, i.e. 
\begin{equation}\label{stability}
    p_0(r)+\frac{2}{3} \,s_0(r) > 0.
\end{equation}
Thus, the positive $D$ term in our approach may imply that $\Omega^-$ is not 
be stable according to the above point of view of Ref.~\cite{Perevalova:2016dln}. 
We claim that we have demonstrated that the positive $D$-term for the $\Delta$ 
resonance in our quark-diquark approach~\cite{Fu:2022rkn}. Note that the similar 
positive $D$-term is also obtained in the calculation of hydrogen atom in 
Ref.~\cite{Ji:2021mfb}. Although our result does not satisfy the inequality 
of~\eqref{stability}, the von Laue condition is indeed satisfied
\begin{equation}\label{vonlaue}
    \int^ \infty _0 d r r^2 p_0(r)=0,
\end{equation}
which can be elucidated by Fig.~\ref{Figurep0r}, where the equality between the 
areas of the upper and lower shaded parts is shown.
\\

\begin{figure}[htbp]
\centering
    \includegraphics[height=6cm]{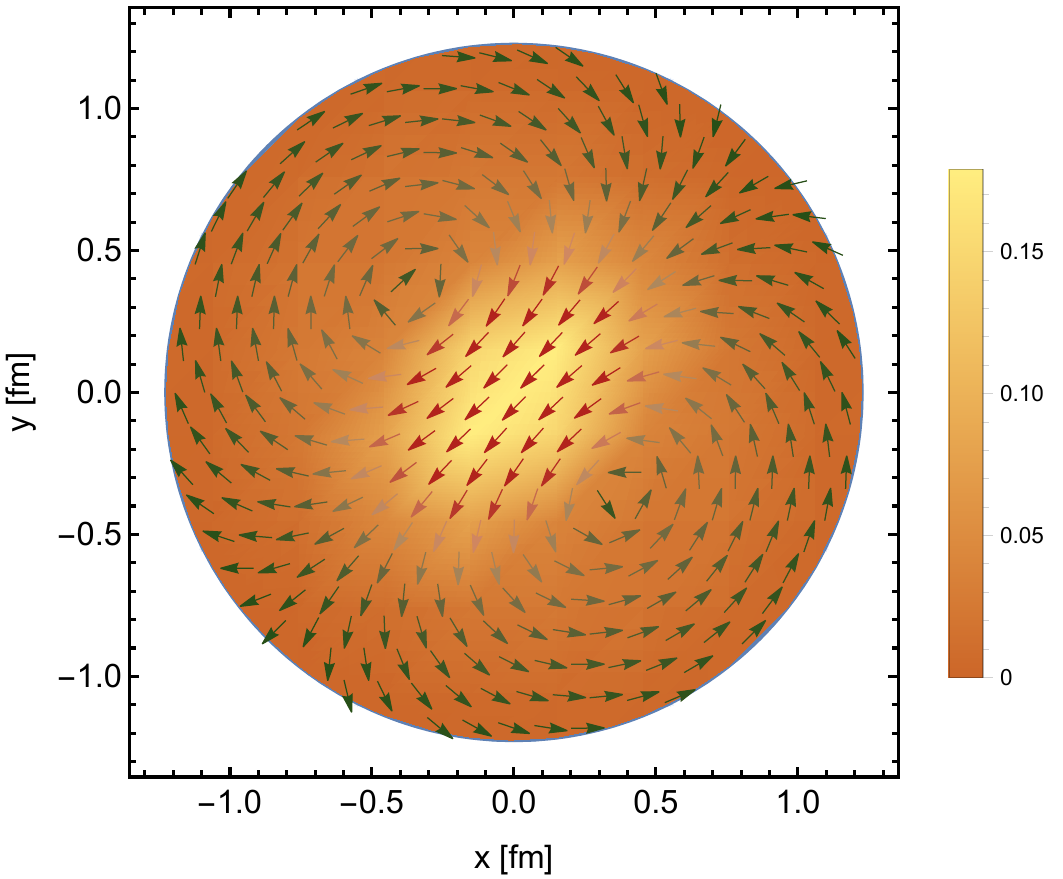}
    \caption{\small{The momentum current with the unit $\text{GeV fm}^{-3}$ on the $x-y$ plane with $z=0$.}}
    \label{FigureMC}
\end{figure}

To explore the $\Omega^-$ stability in more detail, we plot the momentum current 
distribution on the $x-y$ plane with $z=0$ in Fig.~\ref{FigureMC} according to 
Eq.~\eqref{eqTij}, where the arrows and shades represent its direction and strength, 
respectively. It is clearly seen that the absolute value of the momentum current at 
boundary is close to zero. Thus, Fig.~\ref{FigureMC} implies that this is a 
stable system. If we add a minus sign to $D_0(t)$ by hand, the negative $D$-term is 
obtained and only the arrow of each point in Fig.~\ref{FigureMC} points to the 
opposite direction according to Eq.~\eqref{force}, but one can find that the system 
is also stable. Therefore, we argue that the stability is independent of the sign 
of the $D$-term, i.e.  there is no direct relation between the stability of a 
hadron and the sign of the $D$-term as also have been discussed in Ref.~\cite{Ji:2021mfb}.

\section{Summary and discussion}\label{sectionsummary}

\quad\quad In this work, the electromagnetic and gravitational form factors 
of $\Omega^-$ hyperon have been calculated simultaneously using the 
quark-diquark approach. The diquark with two $s$ quarks is considered as an 
axial-vector particle and its specific inner structure is also considered when
we discuss the $\Omega^-$ form factors. In our calculation, we use the 
effective vertex between the hadron and two effective partons, quark and diquark, 
and the simple regularization is also employed to make the integral convergence. 
The model parameters are determined by fitting our EMFFs to the LQCD results.

Our obtained electromagnetic properties of the $\Omega^-$ hyperon, such as its 
magnetic moment, electric-quadrupole moment, electromagnetic radii, and so on, 
are in a reasonable agreement with those of the experiments, LQCD calculations, 
and other models. In addition, we find that the mass radius is smaller than the 
electromagnetic radii. Compared to the results of the $\Delta$ resonance, we 
conclude that $\Omega^-$ has the smaller electromagnetic and mass radii due to the 
stronger boundary. Because of the similar quark components, the behaviours of 
$\Omega^-$ and $\Delta^{++}$ form factors are similar for the low-order ones and 
the energy distribution takes the same prolate shape which can also be illustrated  
by the positive $\varepsilon_2(0)$.

The energy density, angular momentum density, and internal forces including 
pressures and shear forces are also given in the coordinate space by the 
Fourier transformed form factors. An important and fundamental property of the 
system is its stability, and Ref.~\cite{Polyakov:2018zvc} argued that the 
stable system must have the negative $D$-term. However, this explanation is 
still controversial~\cite{Polyakov:2018zvc,Ji:2021mfb,Kim:2022wkc}.
According to our calculations and analyses of GFFs, we believe that there are three 
important issues that need to be stressed and studied.

\begin{enumerate}
   
\item 
There should be mechanical stability and decay stability. The former is due to the 
resultant force being zero at any point and represents the existence of the particle, 
and the latter is because of the forbiddance of its strong decay and represents, at least 
to some extend, the lifetime of the particle. We believe that it is important to distinguish 
between these two types of the stability. The stability, discussed in Ref.~\cite{Polyakov:2018zvc} 
and the related work including this work, should be mechanical stability. We argue that 
all the existent particles, including the unstable particles, even with strong decay modes 
such as $\Delta$ resonances, need to be mechanically stable during their existence.
    
\item 
In this work, the FFs are calculated under the premise that $\Omega^-$ is a 
three quark bound state. We argue that there might be no classical pressure
and shear force in this few-body and hadronic system as well as in the hydrogen atom 
system~\cite{Ji:2021mfb}, because the pressure and shear force result from the statistical mean 
in the multi-body systems. In our opinion, it is more reasonable to use the momentum current 
$T^{ij}(\bm{r})$ as a criterion to judge the mechanical stability, because the momentum 
current does exist in any kind of system.
    
\item 
Moreover, the FFs describe the static properties of the particle in general, and they must 
give a mechanically stable result if the particle exists. As mentioned above and explicitly 
addressed in Ref.~\cite{Fu:2022rkn} and this work, we obtain the positive $D$-term for the 
spin-3/2 particles of $\Delta$ resonances and $\Omega^-$ hyperon in the covariant 
quark-diquark approach. What's the most important is that the obtained momentum flux on any small 
volume is zero whether the $D$-term is positive or negative according to our analyses. Therefore, 
we conclude that the mechanical stability does not relate to the sign of the $D$-term.

\end{enumerate}

\section*{Acknowledgments}
\quad\quad 
We are grateful to Bao-Dong Sun and Volker Burkert for constructive discussions on the relation between the particle stability and the $D$-term.
This work is supported by the National Natural Science Foundation of China under Grants No. 11975245. This work is  also supported by the Sino-German CRC 110 “Symmetries and the  Emergence of Structure in QCD” project by NSFC under Grant No. 12070131001, the Key Research Program of Frontier Sciences,  CAS, under Grant No. Y7292610K1, and the National Key Research  and Development Program of China under Contracts No. 2020YFA0406300.

\bibliographystyle{unsrt}
\bibliography{references}

\end{document}